\input harvmac
\input mssymb
\input labeldefs.tmp
\writedefs
\overfullrule=0pt

\input epsf
\def\fig#1#2#3{
\xdef#1{\the\figno}
\writedef{#1\leftbracket \the\figno}
\nobreak
\par\begingroup\parindent=0pt\leftskip=1cm\rightskip=1cm\parindent=0pt
\baselineskip=11pt
\midinsert
\centerline{#3}
\vskip 12pt
{\bf Fig. \the\figno:} #2\par
\endinsert\endgroup\par
\goodbreak
\global\advance\figno by1
}
\newwrite\tfile\global\newcount\tabno \global\tabno=1
\def\tab#1#2#3{
\xdef#1{\the\tabno}
\writedef{#1\leftbracket \the\tabno}
\nobreak
\par\begingroup\parindent=0pt\leftskip=1cm\rightskip=1cm\parindent=0pt
\baselineskip=11pt
\midinsert
\centerline{#3}
\vskip 12pt
{\bf Tab. \the\tabno:} #2\par
\endinsert\endgroup\par
\goodbreak
\global\advance\tabno by1
}
\def\der{\partial}
\def\d{{\rm d}}
\def\e#1{{\rm e}^{#1}}
\def\las{\{ \lambda_j\}}
\def\mus{\{ \mu _k \}}
\def\A{{\bf A}}
\def\B{{\bf B}}
\def\C{{\bf C}}
\def\D{{\bf D}}
\def\F{{\bf R}}
\def\T{{\bf T}}
\def\Q{{\rm Q}}
\def\a{{\rm a}}
\def\bra#1{\big< #1 \big|\,}
\def\ket#1{\,\big| #1 \big>}
\def\braket#1#2{\big< #1 \big| #2 \big>}
\def\Im{\mathop{\rm Im}\nolimits} 
\def\one{^{1}}
\def\two{^{2}}
\def\alp{^{\alpha}}
%
\def\pre#1{ (preprint {\tt #1})}
%
%
%
\lref\ICK{A.G.~Izergin, D.A.~Coker and V.E.~Korepin,
{\it J. Phys.} A 25 (1992), 4315.}
\lref\Ize{A.G.~Izergin, {\it Sov. Phys. Dokl.} 32 (1987), 878.}
\lref\Suth{ B. Sutherland, {\it PRL} 19 (1967), 103.}
\lref\Kor{V.E. Korepin, {\it Commun. Math. Phys} 86 (1982), 391.}
\lref\Bax{R.J.~Baxter, {\sl Exactly Solved Models in
Statistical Mechanics} (San Diego, CA: Academic).}
\lref\Kup{G.~Kuperberg, 
{\it Internat. Math. Res. Notices} 3 (1996), 139\pre{math.CO/9712207}.}
\lref\Sog{K.~Sogo, {\it Journ. Phys. Soc. Japan}
62, 6 (1993), 1887.}
\lref\Hir{R. Hirota, {\it Journ. Phys. Soc. Japan} 56 (1987), 4285.}
\lref\Wieg {O. Lipan, P.B. Wiegmann, A. Zabrodin,  preprint {\tt solv-int/9704015}.}
\lref\WZ {P.B. Wiegmann, A. Zabrodin, preprint {\tt hep-th/9909147}.}
\lref\WZK{I. Krichever, O. Lipan, P. Wiegmann and A. Zabrodin, preprint {\tt hep-th/9604080}.}
\lref\UT{K.~Ueno and K.~Takasaki, {\it Adv. Studies in Pure Math.} 4 (1984), 1.}
\lref\AvM{M.~Adler and P.~van~Moerbeke, {\it Duke Math. Journal}
80 (1995), 863\pre{solv-int/9706010};
preprint {\tt math.CO/9912143}.}
\lref\To{M. Toda, {\it Journ. Phys. Soc. Japan} 22 (1967), 431;
{\it Prog. Theor. Phys. Suppl.} 45 (1970), 174.}
\lref\Fla{H. Flaschka, {\it Phys. Rev.} B 9 (1974); {\it Prog. Theor.
Phys.} 51 (1974), 703.}
\lref\GMMMO{A.~Gerasimov, A.~Marshakov, A.~Mironov, A.~Morozov
and A.~Orlov, {\it Nucl. Phys.} B 357 (1991), 565.}
\lref\L{E.~Lieb, {\it Phys. Rev. Lett. } 18 (1967), 692.}
\lref\Li{E.~Lieb, {\it Phys. Rev. Lett.} 18 (1967), 1046.}
\lref\Lie{E.~Lieb, {\it Phys. Rev. Lett.} 19 (1967), 108.}
\lref\Lieb{E.~Lieb,  {\it  Phys. Rev.} 162 (1967), 162.}
\lref\Kas{P.W.~Kasteleyn, {\it Physica} 27 (1960), 1209.}
\lref\Fi{M.E.~Fisher, {\it Phys. Rev.} 124 (1961), 1664.}
\lref\LW{W.T.~Lu and F.Y.~Wu, {\it Phys. lett. A } 259 (1999), 108.}
\lref\JPS{W.~Jockush, J.~Propp and P.~Shor, preprint {\tt math.CO/9801068}.}
%
\lref\CEP{H.~Cohn, N.~Elkies and J.~Propp, {\it Duke Math. Journal}
85 (1996), 117.}
%
\lref\EKLP{N.~Elkies, G.~Kuperberg, M.~Larsen and J.~Propp,
{\it Journal of Algebraic Combinatorics} 1 (1922), 111; 219.}
\lref\Br{D.M.~Bressoud, {\sl Proofs and Confirmations:
The Story of the Alternating Sign Matrix Conjecture}
(Cambridge University Press, 1999)}
\lref\BP{D.~Bressoud and J.~Propp, {\it Notices of the AMS} June/July (1999), 637.}
\lref\Deift{P.~Deift, T.~Kriecherbauer, K.T-R.~McLaughlin, S.~Venakides and 
X.~Zhou, {\it Commun. on Pure and Applied Math.} 52
(1999), 1491.}
\lref\BBOY{M.T.~Batchelor, R.J.~Baxter, M.J.~O'Rourke and C.M.~Yung,
{\it J. Phys.} A28 (1995), 2759.}
\lref\Ken{R.~Kenyon, {\sl The planar dimer model with boundary:
a survey}, preprint\hfil\break
({\tt http://topo.math.u-psud.fr/$\sim$kenyon/papers.html}).}
%
\lref\KZJ{V.~Korepin and P.~Zinn-Justin, preprint {\tt cond-mat/0004250}.}
\lref\Sla{J.C.~Slater, {\it Journ. Chem. Phys.} v 9 (1941), 16.}
\lref\Elo{K.~Eloranta, {\it J. Stat. Phys.}
 Vol 96, 5/6 (1999), 1091.}
\lref\OB{R.J.~Baxter and A.L.~Owczarek, {\it J. Phys.} A22 (1989), 1141.}
\lref\BIK{V.~Korepin, N.~Bogoliubov, A.~Izergin, {\sl 
Quantum Inverse Scattering Method
and Correlation Functions} (Cambridge University Press, 1993).}
\lref\MT{J.M.~Maillet and V.~Terras, {\it Nucl. Phys.} B575 (2000),
627.}
\lref\KMT{N.~Kitanine, J.M.~Maillet and V.~Terras,
preprint {\tt math-ph/9907019}.}
\lref\KIEU{V.~Korepin, A.~Izergin, F.~Essler and D.~Uglov,
{\it Phys. Lett.} A190 (1994), 182.}
\lref\GK{F.~G\"ohmann and V.~Korepin, {\it J. Phys.} A33 (2000), 1.}
\lref\Luk{S.~Lukyanov, preprint {\tt cond-mat/9809254}.}
\lref\Aff{I.~Affleck, preprint {\tt cond-mat/9802045}.}
\lref\JMMN{M.~Jimbo, K.~Miki, T.~Miwa and A.~Nakayashiki,
{\it Phys. Lett.} A166 (1992), 256.}
\lref\DFMC{T.~Deguchi, K.~Fabricius and B.~McCoy, preprint
{\tt cond-mat/9912141}.}
%
\Title{
\vbox{\baselineskip12pt\hbox{YITP-00-48}\hbox{{\tt solv-int/0008030}}}}
{{\vbox {
\vskip-10mm
\centerline{\bf Inhomogeneous Six-Vertex Model}
\vskip2pt
\centerline{\bf with Domain Wall Boundary Conditions}
\vskip2pt
\centerline{\bf and Bethe Ansatz}
}}}
\medskip
\centerline{V.~Korepin {\it and} P.~Zinn-Justin}\medskip
\centerline{\sl C.N.~Yang Institute for Theoretical Physics}
\centerline{\sl State University of New York at Stony Brook}
\centerline{\sl Stony Brook, NY 11794--3840, USA}
\vskip .2in
\noindent 

In this note, we consider the six-vertex model with domain wall boundary conditions, defined on a $M\times M$ lattice, in the inhomogeneous case where
the partition function depends on $2M$ inhomogeneities $\lambda_j$ and
$\mu_k$. For a particular choice of the set of $\lambda_j$ we find
a new determinant representation for the partition function,
which allows evaluation of the bulk free energy in the thermodynamic limit. 
This provides a new connection between two types of determinant formulae.
We also show in a special case that spin correlations
on the horizontal line going through the center coincide with the ones for
periodic boundary conditions.

\Date{08/2000}

\newsec{Introduction}
The six-vertex model was first introduced in \Sla.
It was solved exactly by E. Lieb \Lieb\ 
and B.~Sutherland \Suth\ in 1967 by means of  Bethe Ansatz for periodic
boundary  conditions (PBC).
Later the six-vertex model was studied for different boundary 
conditions \BBOY,\OB,\Elo.
Domain wall boundary conditions were introduced in 1982 \Kor. 
These boundary conditions are interesting
because they allow to derive determinant representations for correlation 
functions \BIK\ and the same
boundary conditions help to enumerate alternating sign
matrices \Kup, \Br.
Recently the bulk free energy was 
calculated for these boundary conditions \KZJ.

In this paper we show that for special choices of inhomogeneities,
one can compute the free energy and some
correlation functions of the system. This observation might be useful
because we expect some properties of the model to be independent of
the inhomogeneities i.e.\ to depend only on the anisotropy parameter.
In the simplest situation, the correlation functions
coincide with the ones for periodic boundary conditions.

\newsec{Inhomogeneities and Bethe Ansatz}
In this section we define the inhomogeneous
six-vertex model with domain wall boundary conditions
and choose the spectral parameters
(inhomogeneities) to satisfy Bethe Ansatz equations; this will imply
special properties of the partition function.

We now introduce the notations.
Given a $M\times M$ square lattice with spectral parameters
$\lambda_i$ and $\mu_k$ attached to the lines and columns, one defines
the usual Boltzmann weights $a$, $b$, $c$ to be
\eqn\weib{
\eqalign{
a(\lambda,\mu )&=\sinh(\gamma(\lambda-\mu +i/2))\cr
b(\lambda,\mu )&=\sinh(\gamma(\lambda-\mu -i/2))\cr
c(\lambda,\mu )&=\sinh(i\gamma)\cr
}}
where $\gamma$ is the anisotropy. 
We have {\it fixed} boundary conditions
for the external edges: horizontal (resp.\ vertical)
external edges are outgoing (resp.\ incoming).
The partition function is denoted
by $Z_M(\las,\mus)$. Using recursion relations satisfied by the $Z_M$,
one can show that the following determinant formula holds \Ize, \ICK:
\eqn\detrep{\eqalign{
Z_M(\las,\mus)=&{\prod_{1\le j,k\le M}\sinh(\gamma(\lambda_j-\mu _k+i/2))
\sinh(\gamma(\lambda_j-\mu _k-i/2))
\over\prod_{1\le j<j'\le M} \sinh(\gamma(\lambda_j-\lambda_{j'}))
\prod_{1\le k<k'\le M} \sinh(\gamma(\mu _k-\mu _{k'}))}\cr
&\det_{1\le j,k\le M}
\left[{\sinh(i\gamma)\over\sinh(\gamma(\lambda_j-\mu _k+i/2))
\sinh(\gamma(\lambda_j-\mu _k-i/2))}\right]\cr}
}
In the homogeneous case this represention was used in order to evaluate
the bulk free energy in the thermodynamic limit \KZJ. In a special
inhomogeneous case we shall use another determinant represention
to evaluate the bulk free energy in the thermodynamic limit.
Let us define our special case.
We  choose the spectral parameters in relation to the Bethe Ansatz.
In order to do that it is convenient to introduce the formalism
of the Algebraic Bethe Ansatz. The Boltmann weights are encoded into
the {\it $L$-matrix}
\eqn\Lmat{L(\lambda)=\pmatrix{
a(\lambda)&0&0&0\cr
0&b(\lambda)&c(\lambda)&0\cr
0&c(\lambda)&b(\lambda)&0\cr
0&0&0&a(\lambda)\cr
}
}
We next introduce the {\it monodromy matrix} $T(\lambda)$ which is
an operator acting on ${\Bbb C}^{2^M}\otimes {\Bbb C}^2$ 
(physical space times auxiliary space) defined by
\eqn\Tmat{
T(\lambda;\mu_1,\ldots,\mu_M)=L(\lambda-\mu_M)L(\lambda-\mu_{M-1})\ldots L(\lambda-\mu_1)
}
where $L(\lambda-\mu_k)$ acts on the $k^{\rm th}$ factor of the tensor product
in the physical space, and the auxiliary space.
As an operator on the two-dimensional auxiliary space, $T(\lambda)$
can be written as
\eqn\Tmatb{
T(\lambda)=\pmatrix{\A(\lambda)&\B(\lambda)\cr \C(\lambda)&\D(\lambda)\cr}
}
where $\A$, $\B$, $\C$, $\D$ are operators on the
physical space ${\Bbb C}^{2^M}$.
The usual {\it transfer matrix} corresponding to periodic boundary conditions
is defined as 
\eqn\Tmatc{
\T(\lambda)=\A(\lambda)+\D(\lambda)
}
We shall make use of this operator later. Here, our fixed boundary 
conditions imply the following formal expression for the partition function
\Kor:
\eqn\ABA{
Z_M(\las,\mus)=\bra{\downarrow}
\B(\lambda_1;\mu_1,\ldots,\mu_M)\ldots \B(\lambda_M;\mu_1,\ldots,\mu_M)
\ket{\uparrow}
}
where $\ket{\uparrow}\equiv\left({1\atop0}\right)^{\otimes M}$ (resp.\ 
$\ket{\downarrow}\equiv\left({0\atop1}\right)^{\otimes M}$) 
is the state with all spins up (resp.\ down).

Let us specify inhomogeneities.
We choose the $\las$ to be divided into two sets
$\{ \lambda\one_j=\lambda_j,\, j=1,\ldots, m\one\}$
and 
$\{ \lambda\two_j=\lambda_{j+m\one},\, j=1,\ldots, m\two\}$
which each satisfy the {\it Bethe Ansatz Equations}:
\eqn\bae{
\prod_{\scriptstyle j'=1\atop\scriptstyle j'\ne j}^{m\alp} 
{\sinh(\gamma(\lambda\alp_j-\lambda\alp_{j'}+i))
\over \sinh(\gamma(\lambda\alp_j-\lambda\alp_{j'}-i))}
=\prod_{k=1}^M
{\sinh(\gamma(\lambda\alp_j-\mu_k+i/2))
\over \sinh(\gamma(\lambda\alp_j-\mu_k-i/2))}
}
with $\alpha=1,2$.
We define the left and right states
\eqn\bethe{
\eqalign{
\bra{1}
&=\bra{\uparrow}\C(\lambda\one_1)\ldots \C(\lambda\one_{m\one})\cr
\ket{2}
&=\B(\lambda\two_1)\ldots \B(\lambda\two_{m\two})\ket{\uparrow}
\cr}
}
and the {\it flip operator} $\F$ to be the operator on the physical space
that flips all arrows: in terms of the usual Pauli matrices,
$\F=\prod_{k=1}^M \sigma^x_k$. Now using the overall invariance of
the monodromy matrix under flip: $[T(\lambda),\F\sigma^x]=0$ (where
the additional $\sigma^x$ acts on the auxiliary space), one finds
that $\F\, \B(\lambda)\, \F= \C(\lambda)$ and therefore we can rewrite formula
\ABA\ in terms of the states we have defined:
\eqn\ABAb{
Z_M(\las,\mus)=\bra{1}\F\ket{2}
}

We now consider the situation where $M$ is even and $m\one=m\two=M/2$.
As proven in the appendix ({\bf A.3}),
the Bethe state $\ket{2}$ is an eigenstate of $\F$ 
(with eigenvalue $\pm 1$). At this point we use orthogonality
of Bethe states ({\bf A.2}) to conclude that
\eqn\ABAc{
Z_M(\las,\mus)=\pm\delta_{
\{ \lambda\one_j\},
\{ \lambda\two_j\}
}
\braket{1}{2}
}
The non-zero scalar product (square of the norm) is given by
the following formula \Kor, dropping the superscripts:
\eqn\ABAd{
\braket{1}{1}=(\sin \gamma)^{M/2}
\left[\prod_{j=1}^{M/2} \a(\lambda_j) \d(\lambda_j)\right]
\left[\prod_{\scriptstyle j,j'=1\atop\scriptstyle j\ne j'}^{M/2}
{\sinh(\gamma(\lambda_j-\lambda_{j'}+i))\over\sinh(\gamma(\lambda_j-\lambda_{j'}))}
\right]
\det_{1\le j,j'\le M/2} \left[{\der\varphi_j\over\der\lambda_{j'}}\right]
}
with the following definitions:
$\a(\lambda)$ and $\d(\lambda)$ are the eigenvalues of $\A(\lambda)$
and $\D(\lambda)$ acting on $\ket{\uparrow}$:
\eqn\eigad{
\eqalign{
\a(\lambda)&=\prod_{k=1}^M \sinh(\gamma(\lambda-\mu_k+i/2))\cr
\d(\lambda)&=\prod_{k=1}^M \sinh(\gamma(\lambda-\mu_k-i/2))\cr
}}
and the $\varphi_j$ are the logarithms of the B.A.E.\ \bae:
\eqn\defphi{
\varphi_j=i\log(\a(\lambda_j)/\d(\lambda_j))+i\sum_{\scriptstyle j'=1
\atop\scriptstyle j'\ne j}^{M/2} \log{\sinh(\gamma(\lambda_j-\lambda_{j'}+i))
\over\sinh(\gamma(\lambda_j-\lambda_{j'}-i))}
}
Note that the general determinant formula \detrep, in the case of 2
identical sets $\{ \lambda\one_j\}=\{ \lambda\two_j\}$, becomes
\eqnn\detrepb
$$
\eqalignno{
Z_M(\las,\mus)=&{\prod_{1\le j\le M/2\atop 1\le k\le M}
\sinh^2(\gamma(\lambda_j-\mu _k+i/2))
\sinh^2(\gamma(\lambda_j-\mu _k-i/2))
\over\prod_{1\le j<j'\le M/2} \sinh^2(\gamma(\lambda_j-\lambda_{j'}))
\prod_{1\le k<k'\le M} \sinh(\gamma(\mu _k-\mu _{k'}))}&\cr
&\det_{\scriptstyle 1\le j\le M/2\atop\scriptstyle 1\le k\le M}
\left[\phi(\lambda_j-\mu_k),\psi(\lambda_j-\mu_k)\right]&\detrepb\cr}
$$
with $\phi(\lambda)\equiv \sinh(i\gamma)/
(\sinh(\gamma(\lambda+i/2))\sinh(\gamma(\lambda-i/2)))$ and
$\psi(\lambda)={1\over\gamma}{\d\over\d\lambda}\phi(\lambda)$.
Our new determinant representation \ABAd\ is quite different from 
Eq.~\detrepb;
it is in particular much easier to study its thermodynamic 
limit.

\newsec{Correlation functions}
We have studied so far the function partition of the system; what about correlation functions?

It is unfortunately not possible to find such a simple expression
for an arbitrary correlation function; however, if we restrict
ourselves to the case 
of special correlations which lie on a fixed horizontal line,
then one can again derive determinant formulae for them.

To be specific, 
let us assume again that
the $\las$ consist of two identical sets $\{ \lambda_j,\,j=1\ldots M/2\}$.
Then the probability that all arrows
located at columns $k_1,\ldots,k_n$ and between lines $M/2$ and
$M/2+1$ are up, is given by
\eqn\avg{\left< \pi_{k_1}\ldots \pi_{k_n}\right>\equiv{
\bra{\downarrow}
\B(\lambda_1)\ldots \B(\lambda_{M/2})\, \pi_{k_1}\ldots\pi_{k_n}\,
\B(\lambda_1)\ldots \B(\lambda_{M/2})
\ket{\uparrow}
\over
\bra{\downarrow}
\B(\lambda_1)\ldots \B(\lambda_{M/2})\,
\B(\lambda_1)\ldots \B(\lambda_{M/2})
\ket{\uparrow}
}
}
where $\pi_k\equiv {1\over2}
(1+\sigma^z_k)=\left({1\atop0}{0\atop0}\right)$
acts on the $k^{\rm th}$ space.

Similarly to what was done for the partition function, one can transform
Eq.~\avg\ using the flip operator $\F$ and find:
\eqn\avgb{
\left< \pi_{k_1}\ldots \pi_{k_n}\right>=
{\bra{1}\pi_{k_1}\ldots\pi_{k_n}\ket{1}
\over
\braket{1}{1}}
}
In other words this is simply the usual correlation functions
of the spin operators ${1\over2}(1+\sigma^z)$
for the corresponding spin chain. The computation of these averages is a 
well-known problem, and the general strategy to perform it is by now well 
understood \MT. One must use the solution of the quantum
inverse scattering problem for these operators \refs{\KMT,\GK}:
\eqn\qism{
\pi_k=\prod_{l=1}^{k-1} \T(\mu_l+i/2)\, \A(\mu_k+i/2)\,
\prod_{l=k+1}^M \T(\mu_l+i/2)
}
and then use the fact that our state $\ket{1}$ is an eigenstate of the
$\T(\lambda)$, as well as the commutation relations satisfied by
the $\A(\lambda)$ and $\B(\lambda)$.
In this way we can reduce Eq.~\avg\ (resp.\ Eq.~\avgb) to a sum of expressions
of the type $\bra{\downarrow} \B(\tilde{\lambda}_1)\ldots
\B(\tilde{\lambda}_{M/2})\B(\lambda_1)\ldots \B(\lambda_{M/2})\ket{\uparrow}$
(resp.\ $\bra{\uparrow}\C(\tilde{\lambda}_1)\ldots
\C(\tilde{\lambda}_{M/2})\B(\lambda_1)\ldots \B(\lambda_{M/2})\ket{\uparrow}$).
These can finally be expressed as determinants
either using Eq.~\detrep, or according to the general formula for
scalar products (p.~237 of \BIK).

The case $n=1$ is trivial: $\ket{1}$ is an eigenstate of $\F$, and therefore
$\left<\pi_k\right>=1/2$. The simplest higher correlation function is
the emptiness formation probability where all $k_i$ are 
nearest neighbors, in which
case we obtain immediately
\eqn\empt{
\left<\pi_{k+1}\ldots\pi_{k+n}\right>=
\prod_{l=k+1}^{k+n} \left[\a(\mu_l+i/2)\prod_{j=1}^{M/2}
{\sinh(\gamma(\mu_l-\lambda_j+i/2))\over\sinh(\gamma(\mu_l-\lambda_j-i/2))}
\right]
\bra{1} \prod_{l=k+1}^{k+n} \A(\mu_l+i/2)\ket{1}
}
We shall give the thermodynamic limit of this expression in a particular case.

\newsec{Thermodynamic limit}
We shall show in a particular example how to take the thermodynamic limit
in formulae \ABAd\ and \empt.
We set all $\mu_k$ to $0$ and consider the critical regime
i.e.\ $\gamma$ real. We specify the state $\ket{1}$ to be the ground
state of the transfer matrices $\T(\lambda)$ (or of the Hamiltonian
of the corresponding $XXZ$ spin chain). In the limit $M\to\infty$ the
$\lambda_j$ form a continuous distribution on the real axis determined
by its density $\rho(\lambda)=1/(2\cosh(\pi\lambda))$.

Let us first consider the free energy. One can show that the undetermined
sign in Eq.~\ABAc\ is $+$, and therefore we have
$F=-\log\braket{1}{1}$. We now analyze Eq.~\ABAd\ in the large $M$ limit.
We see that we have
\eqnn\free
$$
\eqalignno{
-F\approx M^2\bigg[
&{1\over2}\int\d\lambda\rho(\lambda)
\log[\sinh(\gamma(\lambda-i/2))\sinh(\gamma(\lambda+i/2))]&\free\cr
+
&{1\over4}
\int\d\lambda\rho(\lambda)\d\lambda'\rho(\lambda')
\log{\sinh(\gamma(\lambda-\lambda'+i))\over\sinh(\gamma(\lambda-\lambda'))}
\bigg]
+M{1\over2}\log\sin\gamma
+\log\det D&\cr
}
$$
The terms of order $M^2$
form the bulk free energy. In order to go further we have to analyze the
behavior of the determinant. It is easy to see that the determinant of
the matrix $D\equiv\left[{\der\varphi_j\over\der\lambda_{j'}}\right]$
is dominated by its diagonal elements; the latter are, by definition
of $\rho(\lambda)$,
\eqn\diag{
{\der\varphi_j\over\der\lambda_j}=2\pi {M\over 2}\rho(\lambda_j)
}
and therefore
\eqn\diagdet{
\log\det D
\approx 
{M\over 2} \log(\pi M)+
{M\over 2} \int \d\lambda \rho(\lambda) \log \rho(\lambda)
}
This gives us the expansion of the free energy up to linear terms in the size
$M$. Note that similar expressions (with different densities $\rho(\lambda)$)
can be found for other states, as long as they have a proper $M\to\infty$ limit.

As to the correlation functions,
it is known that there is a general multiple integral
representation for correlation functions in the thermodynamic
limit \JMMN\ (see also \Luk, \Aff). We shall not repeat the
derivation here; let us simply mention that
starting for example from \empt, one can prove the following formula
\refs{\KMT,\KIEU}:
\eqnn\emptb
$$
\eqalignno{
\left<\pi_{k+1}\ldots\pi_{k+n}\right>=
2^{-n} \left({\pi\over\zeta}\right)^{n(n-1)/2} \int_{-\infty}^{+\infty}
&\d\rho_1\ldots\d\rho_n \prod_{j<k} {\sinh \pi(\rho_j-\rho_k)
\over\sinh \gamma(\rho_j-\rho_k-i)}&\emptb\cr
&\prod_{j=1}^n {\sinh^{j-1}\gamma(\rho_j-i/2)\sinh^{m-j} \gamma(\rho_j+i/2)
\over\cosh^m \pi\rho_j}&\cr
}
$$
This result is identical to the correlation functions
of the 6-vertex model with periodic boundary conditions.

\appendix{A}{A non-degeneracy property}
We assume in this appendix that $q\equiv\e{i\gamma}$ is generic (i.e.\ not a root of unity, except the isotropic case $q=-1$).
We also assume that the spectral parameters $\mus$ do not form any ``strings''
(i.e.\ $\Im(\mu_k-\mu_l)\ne 1\quad\forall k,l$).
We consider two Bethe states $\ket{1}$ and $\ket{2}$ 
characterized by two sets
$\{ \lambda\one_j,\, j=1,\ldots, m\one\}$
and 
$\{ \lambda\two_j,\, j=1,\ldots, m\two\}$.
Bethe states are eigenstates
of the set of commuting transfer matrices $\T(\lambda)$, with corresponding
eigenvalue
\eqn\eig{
\T(\lambda)\ket{\alpha}=
\left[{\Q\alp(\lambda-i)\over \Q\alp(\lambda)} \a(\lambda)
+{\Q\alp(\lambda+i)\over \Q\alp(\lambda)} \d(\lambda)\right]
\ket{\alpha}
}
where $\a(\lambda)$ and $\d(\lambda)$ are given by Eq.~\eigad\ and
are independent of the state, whereas $\Q\alp$ characterizes 
the $\{ \lambda\alp_j \}$:
\eqn\defQ{
\Q\alp(\lambda)=\prod_{j=1}^{m\alp} \sinh(\gamma(\lambda-\lambda\alp_j))
}
Note that the Bethe Ansatz Equations are simply the equations which
ensure pole cancellation in the eigenvalue of $\T(\lambda)$:
${\rm Res}\ \T(\lambda)\ket{\alpha}_{|\lambda=\lambda\alp_j}=0$.

Because of the symmetry of the transfer matrices $\T(\lambda)$
under the flip operator $\F$, we are only considering states with
$m\alp\le M/2$.

We now assume that $\ket{1}$ and $\ket{2}$ have the
same eigenvalue, that is
\eqn\dege{
{\Q\one(\lambda+i)\over \Q\one(\lambda)} \a(\lambda)
+{\Q\one(\lambda-i)\over \Q\one(\lambda)} \d(\lambda)
=
{\Q\two(\lambda+i)\over \Q\two(\lambda)} \a(\lambda)
+{\Q\two(\lambda-i)\over \Q\two(\lambda)} \d(\lambda)
\qquad \forall \lambda
}
We rewrite this as
\eqn\degeb{
\a(\lambda)\left[ \Q\one(\lambda+i)\Q\two(\lambda)
-\Q\two(\lambda+i)\Q\one(\lambda)\right]
=\d(\lambda)\left[\Q\two(\lambda-i)\Q\one(\lambda)
-\Q\one(\lambda-i)\Q\two(\lambda)\right]
}
Up to an overall prefactor $\e{-2(M+m\one+m\two)\gamma\lambda}$,
both left and right hand sides
are polynomials in $\e{2\gamma\lambda}$ of degree at most $M+m\one+m\two$.
Furthermore they have the following $2M$ known zeroes: $\lambda=\mu_k\pm i/2$,
$k=1,\ldots,M$. If some $\mu_k$ coincide the zeroes have a multiplicity;
however note that a $\mu_k+i/2$ cannot coincide with a $\mu_l-i/2$ (since
the $\mu_k$ are not allowed to form strings). There are now two situations:

\noindent 1) $m\one+m\two<M$. In this case we conclude directly that
both sides of Eq.~\degeb\ are zero.

\noindent 2) $m\one+m\two=M$. Since $m\one\le M/2$ and $m\two\le M/2$,
this can only happen if $m\one=m\two=M/2$. However in this case direct
computation of the highest degree terms of the polynomials in Eq.~\degeb\
shows that they are zero, and therefore they are in fact of degree at most
$2M-1$. Again this means that both sides of the equation are zero.

In either case, we finally find
\eqn\findeg{
{\Q\one(\lambda+i)\over \Q\one(\lambda)}={\Q\two(\lambda+i)\over \Q\two(\lambda)}
\qquad\forall\lambda
}
If $q$ is not a root of unity, this implies immediately that
$\{\lambda\one_j\}=\{\lambda\two_j\}$.
What we have proven is the following result:

\noindent {\bf A.1}. Two Bethe states with $m\le M/2$ (i.e.\ $S^z\ge 0$)
correspond to the same eigenvalues of the $\T(\lambda)$ (for all $\lambda$)
if and only if they are identical.

This has the following two immediate corollaries:

\noindent {\bf A.2}. Two distinct Bethe states are orthogonal to each other.

\noindent {\bf A.3}. A Bethe state with $m=M/2$ (i.e.\ $S^z=0$) is an
eigenstate of the flip operator $\F$.

{\it Remark:} the situation is much more subtle 
if $q$ is a root of unity. One way to see this is to consider Eq.~\findeg,
and assume now that $q^{2N}=1$. One can add an extra ``full string'' of the form
$\lambda_i=\lambda_0+ij$, $j=1,\ldots, N$ to one of the sets without modifying
the corresponding $\Q$ function. This suggests extra degeneracy can appear when
$q$ is a root of unity between states with $\Delta S^z=N$, which is 
precisely the phenomenon observed in \DFMC.
\listrefs
\bye